\documentclass[5p,twocolumn,times,number]{elsarticle}

\usepackage{graphicx}
\usepackage{amsmath}   
\usepackage{hyperref}

\begin{document}

\begin{frontmatter}

\title{Cryogenic light detectors with enhanced performance for rare event physics}

\author[add0]{M.~Barucci}
\author[add1]{J.W.~Beeman}
\author[add5]{V.~Caracciolo}
\author[add2,add3]{L.~Pagnanini}
\author[add4,add5]{L.~Pattavina\corref{tum}}
\author[add3]{G.~Pessina}
\author[add5]{S.~Pirro\corref{cor}}
\ead{Stefano.Pirro@LNGS.INFN.IT}
\author[add6,add5]{C.~Rusconi}
\author[add4,add5]{K.~Sch\"affner}
\cortext[tum]{presently at Technische Universit\"at M\"unchen}
\cortext[cor]{Corresponding author}
\address[add0]{CNR - Istituto Nazionale di Ottica, I-50125, Firenze (FI) - Italy}
\address[add1]{Lawrence Berkeley National Laboratory, Berkeley, California 94720, USA}
\address[add5]{INFN  Laboratori Nazionali del Gran Sasso, I-67100 Assergi (AQ) - Italy}
\address[add2]{Dipartimento di Fisica, Universit\`{a} di Milano Bicocca, I-20126 Milano - Italy}
\address[add3]{INFN - Sezione di Milano-Bicocca,  Milano, Italy}
\address[add4]{Gran Sasso Science Institute, 67100, L'Aquila - Italy}
\address[add6]{Department of Physics and Astronomy, University of South Carolina, Columbia, SC 29208 - USA}

\begin{abstract}

We have developed and tested a new way of coupling bolometric light detectors to scintillating crystal
bolometers based upon simply resting the light detector on the crystal surface, held in position only by gravity.
This straightforward  mounting results in three important improvements: (1) it decreases
the amount of non-active materials needed to assemble the detector, (2) it substantially increases
the light collection efficiency by minimizing the light losses induced by the mounting structure, and 
(3) it enhances the thermal signal induced in the light detector thanks to the extremely weak 
thermal link to the thermal bath. 

We tested this new technique with a 16 cm$^2$ Ge light detector with thermistor readout sitting 
on the surface of a large TeO$_2$ bolometer. The light collection efficiency was increased by greater
than 50\% compared to previously tested alternative mountings.
We obtained a baseline energy resolution on the light detector of 20~eV RMS that, together 
with increased light collection, enabled us to obtain the best $\alpha$ vs $\beta/\gamma$ 
discrimination ever obtained with massive TeO$_2$ crystals. 
At the same time we achieved rise and decay times of 0.8 and 1.6 ms, respectively.
This superb performance meets all of the requirements for the CUPID (CUORE Upgrade with Particle 
IDentification) experiment, which is a  1-ton scintillating bolometer follow up to CUORE.

\end{abstract}

\end{frontmatter}
\section{Introduction}
\label{sec:Introduction}
The use of Low Temperature Detectors (LTDs) for sensing  
 X-ray and $\gamma$-ray signals is quite widespread and well 
established~\cite{Ullom-2015}.
LTDs are also widely used in the field of fundamental physics, especially for 
Double Beta Decay (DBD), and Dark Matter (DM) searches~\cite{Pirro-2017}.
In these surveys the need for a hybrid detector, in which an energy 
release can be measured through different mechanisms,  is of primary importance in order 
to distinguish the nature of interacting particles.  For instance hybrid detectors can help identify 
and reject events caused by the natural background. With thermal detectors this can be 
achieved  using scintillating or luminescent crystals. The simultaneous and 
\it independent \rm readout of the heat  and the (escaping) light produced by the 
interaction  reveals the nature of the interacting particles thanks to 
the different  scintillation yields of $n$, $\alpha$ and $\gamma/\beta$ events.
This discrimination technique is presently used for DM 
searches~\cite{CRESST-2016,Angloher:2017sft,Angloher:2016hbv}, 
DBD searches~\cite{CUPID-0-2018,Cupid-Mo-2017,Amore-2017}, and it can be  
also implemented for rare nuclear decays~\cite{Alfa-1,Alfa-2,Pattavina:2018nhk}.

At milli-Kelvin temperatures, the light detectors are usually bolometers 
themselves: a \it dark \rm thin crystal absorbs the photons, producing heat (phonons) that is
measured by a suitable thermometer.   
The main difference among the various Bolometric Light Detector (BLD) instruments currently in use  
is the choice of the thermometer element, e.g. Transition Edge Sensors 
(TES)~\cite{TES_LD_CRESST}, Neutron  Transmutation Doped  (NTD) 
thermistors~\cite{NTD_LD_Lucifer-2013} or Micro Magnetic Calorimeters (MMC)~\cite{MMC_LD-2015}.  

The work presented here was performed within the CUPID 
framework~\cite{CUORE-IHE-2014,CUPID-2015}, the future follow up of 
CUORE~\cite{CUORE-2018} that represents the largest world-wide bolometric  experiment to date. 
The aim was to  develop NTD-based BLDs with improved performance in terms of 
sensitivity, time response and simplified packaging for large arrays. 
Using the tiny Cherenkov light emission of TeO$_2$~\cite{Tabarelli-2010,Enriched-TeO2-Cherenkov-2017}  to 
decrease by two order of magnitude the  $\alpha$-induced background, requires a BLD with a S/N ratio 
of the order of  $\sim$5~\cite{CUPID-2015}:  this corresponds to a RMS baseline resolution of the BLD of the 
order of $\sim$20~eV  being the Cherenkov  light signal of the order of 100 eV.
Actually one can work towards the optimization  of the light collection~\cite{Casali-2017} and/or 
towards the energy resolution of the BLD or -as we made in this work- both.
Additionally, in case of $^{100}$Mo-based compounds, beside the  same need to suppress the surface 
$\alpha$-induced background,  a fast time response of the BLD ($\leq$~1~ms)  is mandatory to suppress
the background  induced the pile-up of the 2$\nu$ DBD~\cite{2_nu-Pile_up-2012}: also in 
this case the S/N ratio will play an important role~\cite{2_nu-Pile_up-2016}.

Our work has therefore focused on two aspects of BLD performance: (1) improving the response of 
the NTD thermometer and (2) increasing the light collection.
While the first aspect is strictly related to a specific technique, the second aspect 
is worthy of additional remarks. The working principle of a BLD is 
irrespective  of the sensor: a  thin  crystal wafer (usually  Si or Ge)  absorbs 
the emitted photons and  converts them into heat. 
Unlike a conventional bolometric approach, we have to avoid the optical coupling 
between crystal and BLD made with optical grease or similar substance since the unavoidable 
heat flow through the optical coupling and the increase of the heat capacity of the system would  reduce the  
independence of the two  detectors, eliminating the possibility of particle 
discrimination afforded by the different scintillation yields.  Therefore the thermal contact  between 
the luminescent crystal and BLD has to be avoided,
especially in the case of extremely low scintillation yields.  This is true for  most 
of the Mo-based compounds~\cite{Cupid-Mo-2017} and, even more importantly in  case of 
Cherenkov signals. A 2615~keV $\gamma$-ray energy release in a CUORE-like TeO$_2$ absorber 
produces a light signal in the BLD on the order of  $\sim$100~eV~\cite{Casali-2017}. 
For this reason the BLD is always facing the scintillating crystal without directly contacting 
it via a coupling medium.
 
In the following section it is shown that  if the BLD is simply resting on the crystal 
surface, held in  position only by gravity, the thermal coupling between the BLD and 
the crystal is almost negligible and the leakage of the BLD thermal signal through the scintillating crystal vanishes. 
This fact can be explained considering the acoustic mismatch described in 
the diffused mismatch model whereby the  heat carriers (phonons) in insulating materials are 
scattered at the interfaces~\cite{Matsumoto-1977,Swartz-1989}. This approach shows that the 
thermal resistance between two dielectric crystals is strongly dependent on the surface 
state, on the different phonon characteristics in the two materials (density and Debye 
temperature), and on the applied force. This latter parameter has a significant effect.  
When two solids are placed in contact with each other, the actual contact area can be much smaller 
than the cross sections involved due to surface irregularities.  By rising the applied force 
between the materials, a plastic or permanent deformation occurs and the "real" contact surface 
area increases. The result of this action is that the thermal conductance of the contact is directly 
proportional to the applied force~\cite{Barucci-2001,Ventura-2008}.

Although such  simple stand will  clearly not produce  a so-called "optical matching," the 
light collection  will be  definitively larger due to geometrical factors~\footnote{For instance if 
the BLD is held  in its own structure, depending on the mounting scheme, there are generally a 
few mm of distance from the BLD to the scintillating crystal. 
This increases the chance for photon escape or absorption by the holding structure rather than the BLD.}.
In addition, removing the BLD mounting structure decreases the presence of materials and surfaces close
to the detector which reduces possible radioactive contamination, a fundamental aspect of dealing with rare event searches.

\section{Bolometric Light Detectors}

Our BLDs are usually constituted by electronic grade undoped Ge wafers, coupled with  Ge NTD  thermistors. 
We started to develop these detectors coupled with several scintillating DBD 
crystals~\cite{PIRRO-2005} and we deeply characterized  their operation and 
performances~\cite{light-detectors-2013} to finally realize  the  LUCIFER~\cite{LUCIFER-2013} experiment, 
which has been renamed CUPID-0~\cite{CUPID-0-detector_2018}.

Each   BLD  of CUPID-0 (totalling 26 detectors) was made by a double side polished electronic 
grade undoped Ge wafer (44.5~mm diameter, 0.17~mm thick). The NTD thermistor, with dimension of 
(2.85~$\times$~2~$\times$~0.5)~mm$^3$, 
is glued through  six small glue dots ($\sim$~0.5~mm diameter, 0.05~mm height) made with 
Araldit\textsuperscript{\textregistered} Rapid glue.
The performance of six of these detectors was evaluated in a dedicated test run~\cite{LUCIFER-2016} and 
the results are summarized in Tab.~\ref{tab-cupid-0-LD}.

To further optimize our BLDs, we  produced a set of devices based on the pioneering work of 
Coron et.al.~\cite{Coron-2004}.  For this study we (1) decreased the heat capacity (size) of the 
thermistor, (2) increased the thermal conductance between the thermistor and the Ge wafer, and (3) 
decreased the  thermal conductance to  the thermal bath. 
With respect to the  thermistor size, we used thermistors with a dimension of
(2.85~$\times$~1~$\times$ 0.4)~mm$^3$, roughly 2.5 times smaller than the CUPID-0 devices. 
We also decided to replace the six glue dots with an uniform glue layer, thus increasing the thermal 
conductance between the thermistors and light-absorbing Ge wafer. 

It should be noted that in our experience the use of glue dots 
instead of a \it more effective \rm  thin gluing layer is  preferred when coupling inherently different 
materials (e.g. TeO$_2$ crystals and Ge thermistors).  The dot approach reduces the mechanical stresses 
induced by  differential thermal contraction of the materials when cooled. 
In such cases, and especially when working with larger-sized thermistors, we sometimes observed cracks on 
the crystal surface after a cooling cycle. This phenomenon is greatly reduced in our case since we glue 
Germanium thermistors to Germanium light absorbers and use smaller thermistors.  Even in this 
case, however, there are some small unavoidable stresses due to misorientation between the thermistor
and absorber crystallographic planes, but we have found that these effects never led to visible cracks.

With respect to the mounting (i.e. the conductance to the thermal bath), there are many ways  
to hold the BLD in place.  In earlier work we  adopted  two~\cite{light-detectors-2013} or 
three~\cite{CUPID-0-detector_2018} small PTFE  clamps that squeeze the edge of the Ge, keeping 
it fixed in a Cu standalone holder. 
PTFE is a  common material also used by other  groups working with NTD 
sensors~\cite{Lumineu-2017} and  with MMC detectors~\cite{MMC_LD-2015}. 
Other clamping schemes and material choices have been demonstrated by the CRESST group.  
These include bronze clamps and Silicon  or  CaWO$_4$-based sticks~\cite{Strauss-2018}. 
The design used in~\cite{Coron-2004}, however, is probably the most complex from a construction point 
of view, using several ultra thin superconductive wires to suspend the Ge wafer from a copper frame to 
produce a negligible thermal link that maximizes the heat flow from the wafer to the NTD.  
\begin{table}[t]
\centering
\caption{Mean performance of six  CUPID-0-like light detectors~\cite{LUCIFER-2016}.
\textbf{R$_{work}$} refers to the resistance of the NTD Ge thermistor in working conditions, 
\textbf{Response} refers to the absolute voltage 
drop (in $\mu$V) produced by an energy release of 1\,keV,  \textbf{Baseline RMS} is the resolution
after signal filtering~\cite{Gatti-1986:1,Alduino-2016:045503}. {\bf$\tau_{r}$} and {\bf$\tau_{d}$} are the 
rise and  decay  times, computed as the time difference  between the 90$\%$ and 10$\%$ of the leading
edge and as the time difference between the 30$\%$ and 90$\%$ of the trailing edge,
respectively. The Bessel cut-off frequency is 200 Hz (see last remarks of 
Sec.~\ref{sec:results}).}
\label{tab-cupid-0-LD}
\begin{tabular}{ccccc}
\hline\noalign{\smallskip}
R$_{work}$      &Response     &Baseline RMS     &$\tau_{r}$       &$\tau_{d}$  \\

[M$\Omega$]      &[$\mu$V/keV]     &[eV]             &[ms]              &[ms]\\
\hline
   0.87          & 1.36            & 43             & 1.77              & 5.06 \\
\hline
\end{tabular}
\end{table} 

We decided to avoid any kind of holding structure whatsoever so we laid the BLD 
directly on the crystal, kept in position only by its weight ($\sim$1.1 g). 
In this configuration the main thermal link between the BLD and the cryostat is represented by 
the thin gold NTD thermistor wires ( 2 $\times$ 15 mm length, 25 $\mu$m diameter).  As mentioned 
above, the expected thermal conductance to the scintillating crystal is negligible.  The  crystal  
chosen for this test was a  (50.5~$\times$~50.5~$\times$~50.5)~mm$^3$ 
TeO$_2$ crystal.  The aim was to test  the new setup with a light signal on the order 
of few tens of eV. The Ge light-absorbing wafer 
belongs to the batch used for CUPID-0, which include a  70~nm SiO anti-reflecting coating~\cite{Mancuso-2014} 
that was deposited on the side that rests on the TeO$_2$ crystal.

\section{Experimental details}
\label{sec:experimental_details}
The TeO$_2$ crystal  was mounted in a similar way as described 
in~\cite{Enriched-TeO2-Cherenkov-2017,Casali-2014} with the only exception that the TeO$_2$ crystal was standing on 
the reflecting foil and both TeO$_2$ and BLD were not equipped with Si heaters.
These heaters were normally glued 
on the bolometer  to inject pulsed thermal signals for gain stabilization.
The TeO$_2$ face supporting the BLD and the opposite one were polished at (nearly)  optical level. 
The remaining four lateral faces were matted in order to increase light collection~\cite{Casali-2017}. 

The  TeO$_2$  crystal is held by four S-shaped PTFE supports that are fixed to Cu columns. 
The PTFE contracts upon cooling, creating a tensioned support that maintains the crystal 
position. 

\begin{figure}[hbt] 
\centering 
\includegraphics[width=0.48\textwidth]{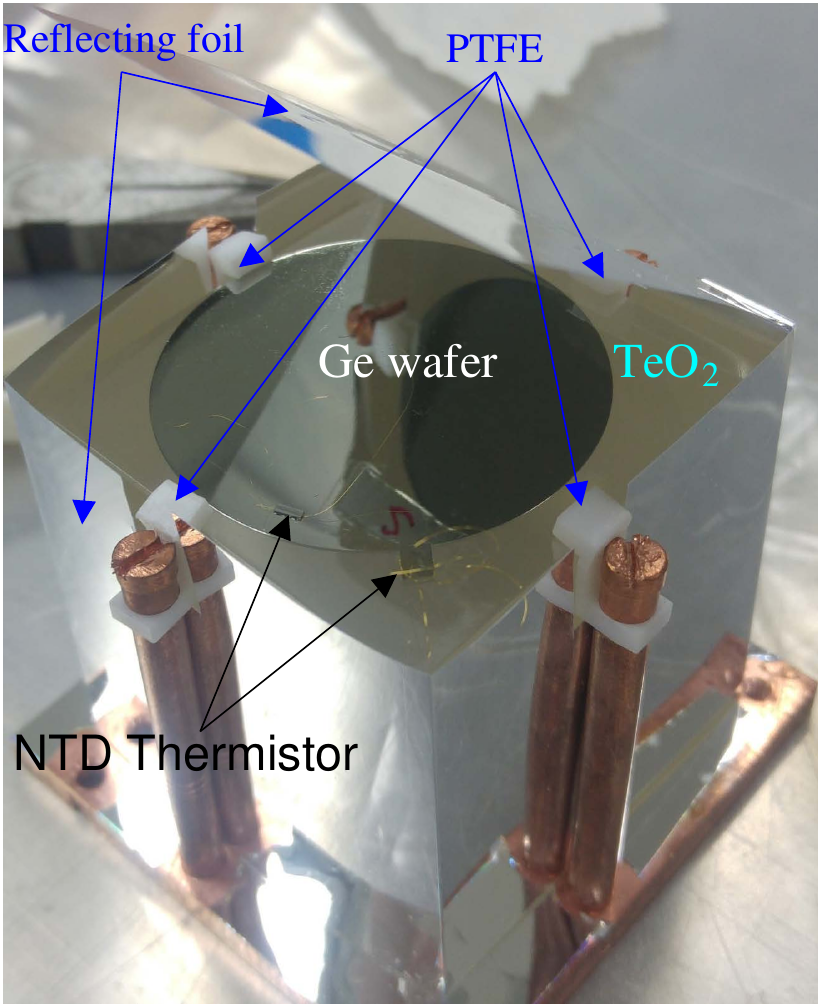}
\caption{Photograph of the detectors. The BLD is simply  resting on the TeO$_2$ and the four 
PTFE supports (as well as the thermistor glued on the  TeO$_2$) do not hold the BLD in any way: they   
simply avoid the BLD to lean out from the top surface, as a mere translation constraints. 
The gold wires of both NTDs are then crimped within micro Cu tubes to ensure the electrical contact 
as well as the thermal conductance to the heat sink. The $^{55}$Fe X-ray source is attached 
to the top reflecting cover sheet that encloses the detectors (with a clearance of $\sim$4~mm from the BLD) 
and can be observed -reflected by the Ge wafer surface- between the two NTDs.}
\label{fig_0-setup}
\end{figure}

In order to maximize light collection, the crystal is completely surrounded by a plastic 
reflecting sheet  (3M Vikuiti\textsuperscript{TM}), in the same way as 
in~\cite{Enriched-TeO2-Cherenkov-2017,Casali-2014}. A photograph of the detectors is presented 
in Fig.~\ref{fig_0-setup}.
The  entire  setup was  enclosed  in a Cu box and thermally  coupled  to the  mixing  chamber  
of the  CUPID R\&D cryostat, a $^3$He/$^4$He  dilution  refrigerator installed  
deep underground  within  Hall  C  of the  Laboratori Nazionali  del Gran Sasso, Italy. 
To  avoid vibrations reaching the detectors, the box is mechanically decoupled from the 
cryostat by utilizing a two-stage pendulum system~\cite{Pirro-2006}.

The thermistors of the detectors are biased  with a quasi-constant current produced by applying a 
fixed voltage through large (27+27 or 2+2 G$\Omega$) load resistors~\cite{Arnaboldi-2002:1808}.   
When light is absorbed in the Ge wafer, a thermal pulse is produced which is subsequently 
transferred to the NTD sensor, changing the resistance of the thermistor.  This, in turn, creates a
voltage change across the current-biased NTD which is amplified using  
front end electronics located just outside the cryostat~\cite{Arnaboldi-2004}. The signals are 
then filtered by an anti-aliasing 6-pole Bessel filter (with a cutoff  frequency of 16~Hz 
for the TeO$_2$ crystal and 550~Hz  for the BLD) and finally fed into a NI PXI-6284 18-bit ADC.

The sampling rate of the ADC was 1~kHz  for the TeO$_2$ crystal and 8 kHz for the BLD. 
The two independent  triggers are software generated such that when a trigger fires, the 
corresponding waveform is recorded. Moreover, when the trigger of the 
TeO$_2$ crystal fires, the corresponding waveform  of the  BLD is always 
recorded, irrespective of its trigger. A detailed description of the DAQ system can 
be found in~\cite{DiDomizio:2018ldc}.
The amplitude and the shape of the voltage pulses are then determined via off-line analysis. 
The pulse amplitude of the thermal signals is estimated by  the Optimum Filtering (OF)  
technique~\cite{Gatti-1986:1,Alduino-2016:045503}, that maximizes the signal-to-noise ratio 
in a way that improves the energy resolution and lowers the threshold of the detector. 
The amplitude of the  light signal, however, is evaluated from the filtered waveform 
at a fixed time delay  with respect to the TeO$_2$ bolometer, as described in detail 
in~\cite{Piperno-2001:10005}.\newline
The amplitude of the acquired TeO$_2$ heat signals is energy-calibrated using several 
$\gamma$-ray peaks from a $^{228}$Th source. 
The BLD, on the contrary, is calibrated thanks to the  5.9~keV  and 6.5~keV X-ray 
quanta  produced by a $^{55}$Fe X-ray source permanently faced  to the detector.

\section{Data analysis and results}
\subsection{BLD performance}
\label{sec:results}
The crystals were  tested at a  cryostat base temperature of $\sim$11~mK. 
In order to obtain a fast response, we operated the BLD in the so-called "over-biased" 
configuration whereby 
the biasing current of the circuit is set much larger than the current that would ensure the highest 
absolute thermal response~\cite{NTD_LD_Lucifer-2013}. This choice ensures a small working 
resistance, thus minimizing the effect of  the low pass filtering  induced by the overall 
capacity  ($\sim$200 pF) of the front end readout wires. 

\begin{figure}[hbt] 
\centering 
\includegraphics[width=0.48\textwidth]{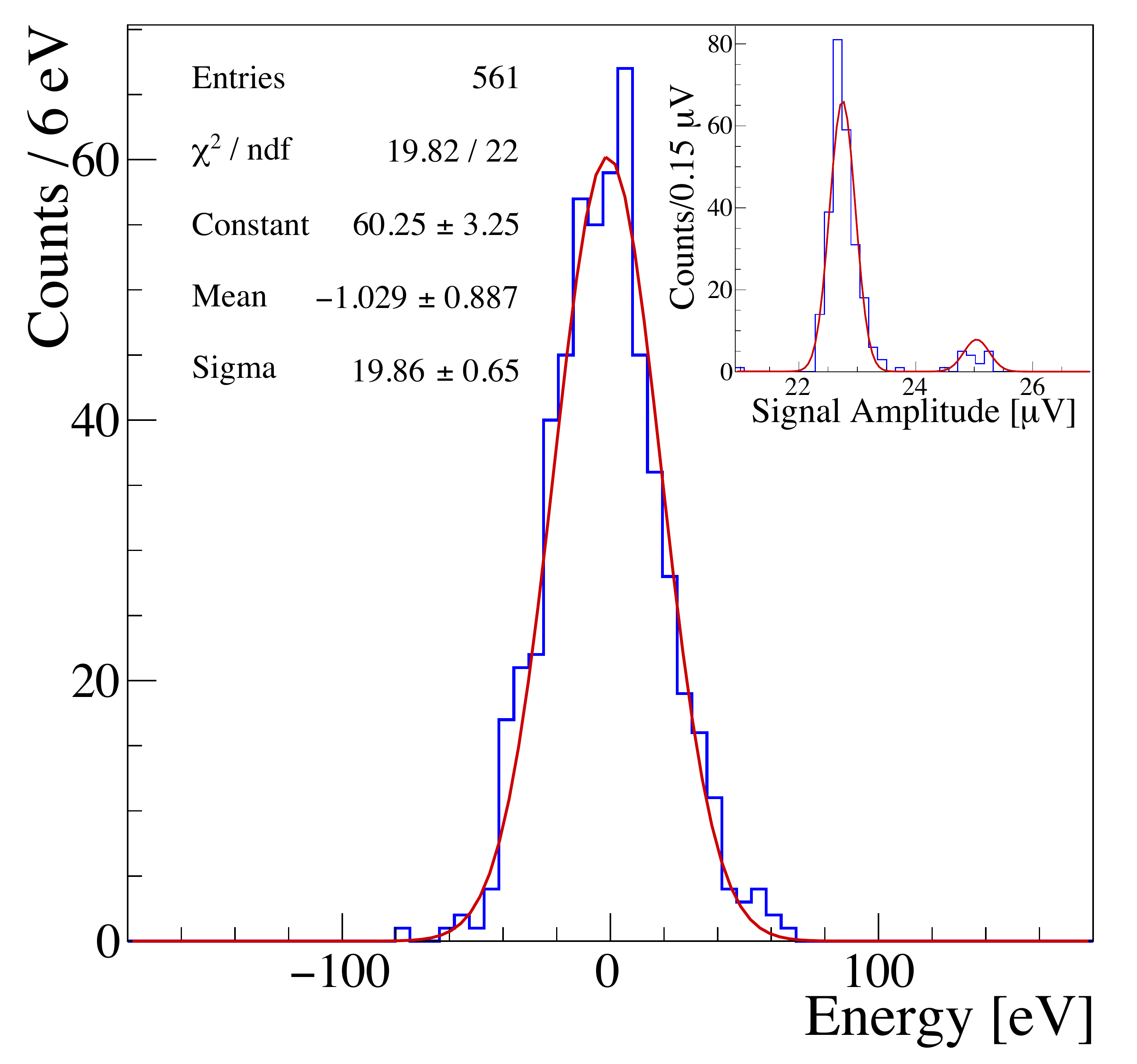}
\caption{Energy distribution of the random sampled noise. The  width of the distribution  
($\sigma\approx$20 eV) represents the baseline energy resolution of our BLD. The right inset shows
the $^{55}$Fe calibration spectrum of the BLD. The x-axis units represent the absolute voltage drop 
across the thermistor. 
The RMS resolution on the 5.9 keV and 6.5 keV X-ray peaks is 59 eV (see text).}
\label{fig_1-55Fe}
\end{figure}

In Fig.~\ref{fig_1-55Fe} we show the $^{55}$Fe calibration spectrum obtained with the BLD.
The baseline energy resolution (ie, the absolute sensitivity) of the BLD is given by 
the width of randomly acquired baselines (noise) after the  application of OF. 
As is typical for this style of detectors, the energy resolution of monochromatic energy 
absorption events is much worse than the baseline resolution, irrespective 
of the type of sensor~\cite{NTD_LD_Lucifer-2013,TES_LD_CRESST}.  

\begin{figure}[hbt] 
\centering 
\includegraphics[width=0.48\textwidth]{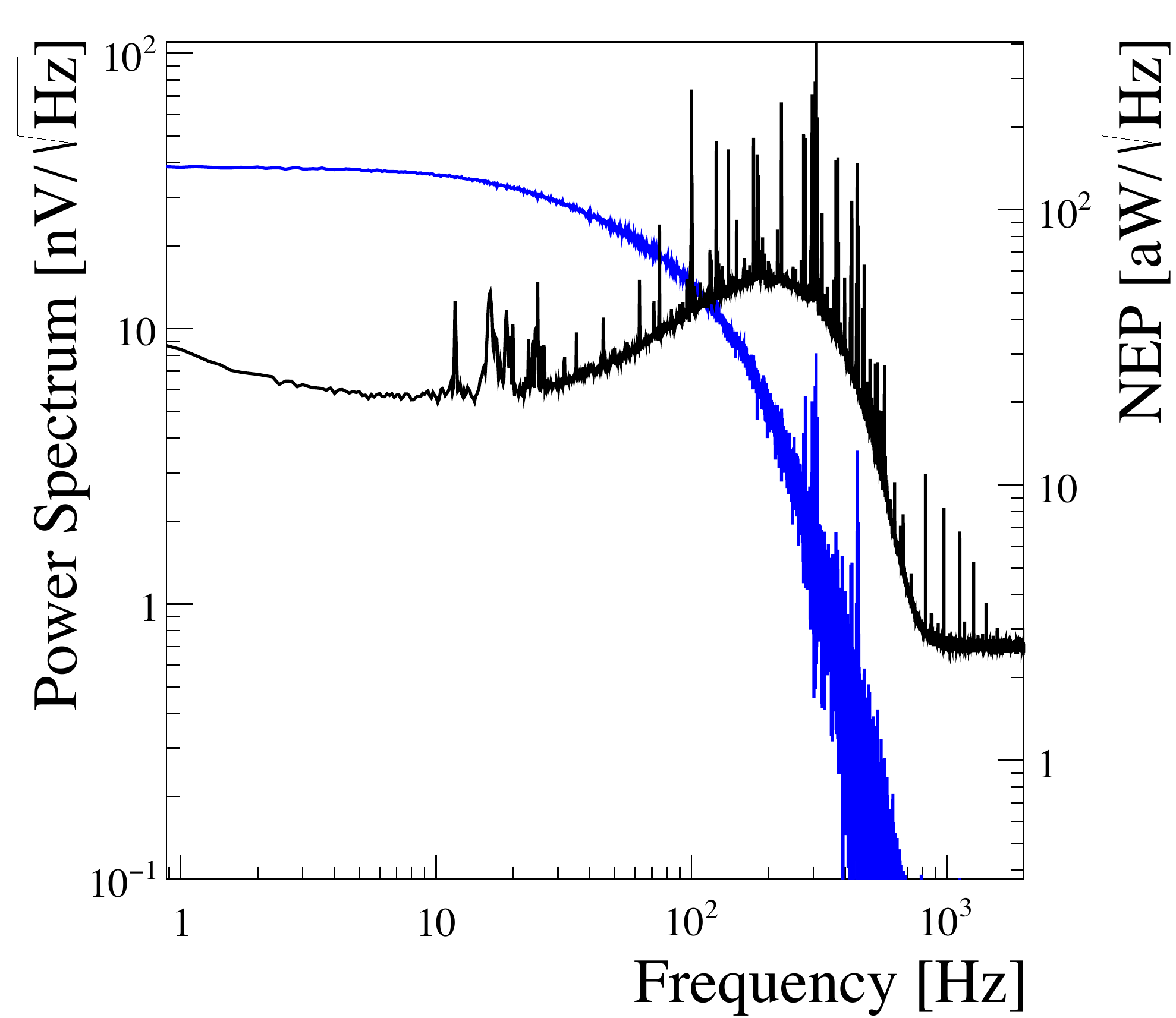}
\caption{Noise power spectrum (black line) and signal power spectrum (blue line) of the BLD. 
The y-axis scale is in absolute values  for the noise. The signal spectrum is scaled  in 
arbitrary units, being the roll-off induced by the Bessel filter  the same between noise and signal.
The working resistance of the thermistor is 1.47 M$\Omega$, biased with a current of 3.7 nA 
thorough (2+2) G$\Omega$ metallic load resistors. The peaks are due to the microphonic 
noise induced by the vibration of the readout wires.}
\label{fig_2-NPS}
\end{figure}

The noise and signal power spectra of the BLD are presented in Fig.~\ref{fig_2-NPS}.

The bump that can be observed in Fig.~\ref{fig_2-NPS} at  $\sim$400 Hz arises from a 
resonance that enhances the thermal noise generated within the thermistor.  This occurs  
when the impedance of the parasitic capacitance of the link becomes smaller than that of the 
thermistor, which is a fed-backed device~\cite{Arnaboldi-2005}. 
The bump is found at the border of  the bandwidth of the signal and is rejected from the 
optimum filter algorithm.

\begin{figure}[hbt] 
\centering 
\includegraphics[width=0.48\textwidth]{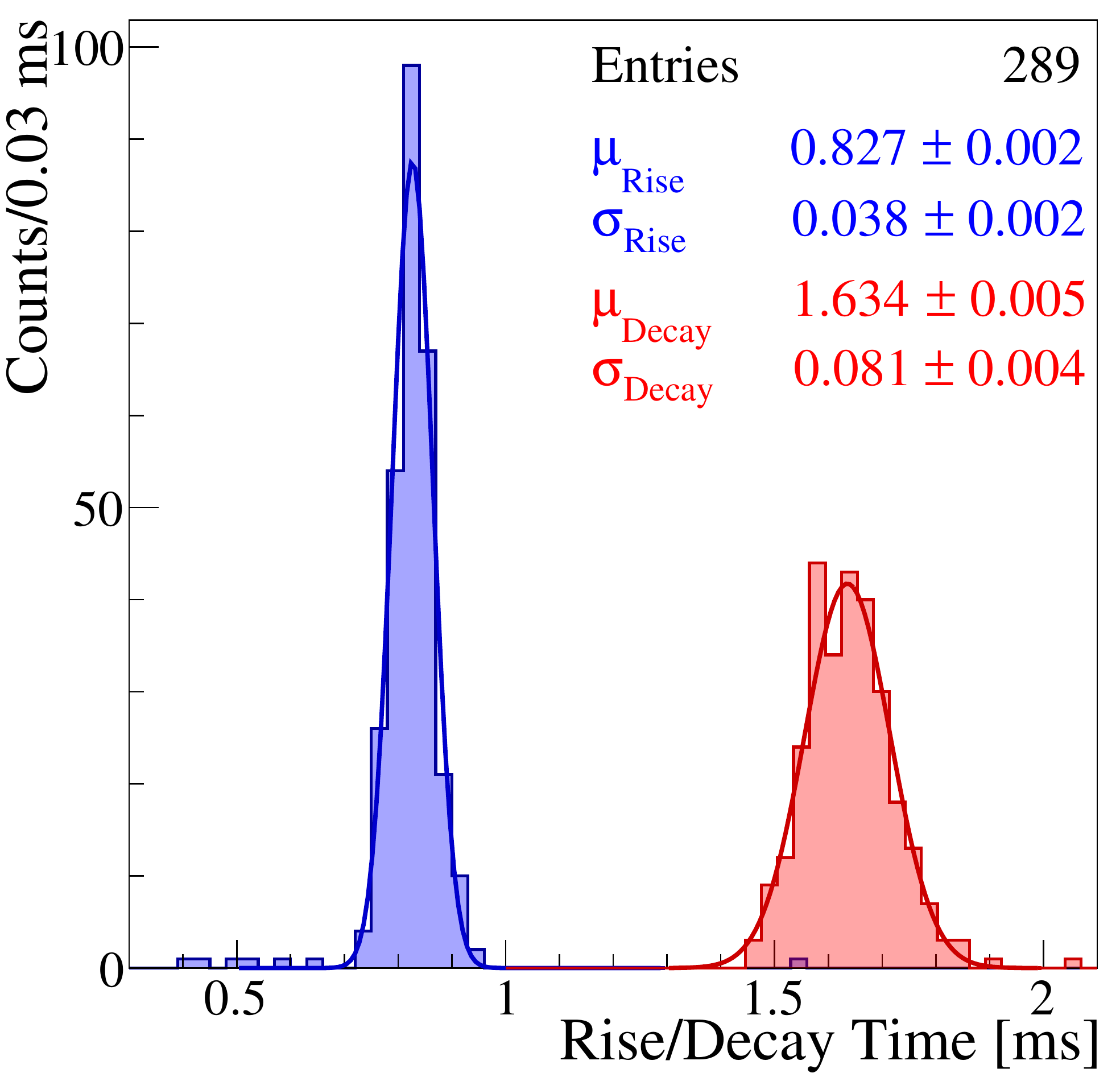}
\caption{Rise and decay times distributions corresponding to the $^{55}$Fe X-rays. The 
Bessel cut-off frequency of the Front-End is 550~Hz.}
\label{fig_3-rise-decay}
\end{figure}

Fig.~\ref{fig_3-rise-decay} shows the corresponding rise and decay times of $^{55}$Fe X-rays absorption events.
The measured rise time shown in Fig.~\ref{fig_3-rise-decay} is most likely slower than the intrinsic 
rise time of the detector since it contains contributions from the Bessel filter (independent from the 
thermistor impedance) and from the capacitance of the readout wires. This last 
contribution is difficult to measure since it involves the dynamic resistance of the 
thermistor. The contribution of the 550~Hz Bessel filter to the rise time was evaluated 
in~\cite{NTD_LD_Lucifer-2013} and reported as 0.65~ms. Thus, after applying a quadratic deconvolution, the 
\it intrinsic \rm rise time of our BLD should be of the order of 0.5~ms, compatible with the 
expectation of ~\cite{Coron-2004}.
The overall performance of the BLD is summarized in Tab.~\ref{tab-new-BLD}.

\begin{table}
\centering
\caption{Performances of the BLD of this work, to be compared with the ones of 
Tab.~\ref{tab-cupid-0-LD}.}
\label{tab-new-BLD}
\begin{tabular}{ccccc}
\hline
R$_{work}$      &Response     &Baseline RMS     &$\tau_{r}$       &$\tau_{d}$ \\

[M$\Omega$]     &[$\mu$V/keV]     &[eV]             &[ms]              &[ms]\\
\hline
1.47             &3.86          &20              & 0.83             &1.63 \\
\hline
\end{tabular}
\end{table} 
\subsection{Heat and Light measurement}
\label{scatter-section}
In order to evaluate the long-term discriminatory performance of our BLD, we performed a 70 h run that 
included two event-generating calibration sources embedded into the setup.  A $^{228}$Th source was placed
a few cm away from the TeO$_2$ crystal and 
a \it smeared \rm $^{238}$U $\alpha$ source was applied to the inside of the light reflector facing the TeO$_2$. 
The aim of the $\alpha$ source was to directly measure the discrimination capability between 
$\alpha$ and $\beta/\gamma$ in the DBD region of interest of $^{130}$Te. 
The source was made using 2 $\mu l$ of a standard calibrated solution (0.1 \%) of $^{238}$U, and the dried 
source deposition was covered with a 6 $\mu m$ aluminized Mylar foil to smear the 
$\alpha$ energy.

\begin{figure}[hbt] 
\centering 
\includegraphics[width=0.48\textwidth]{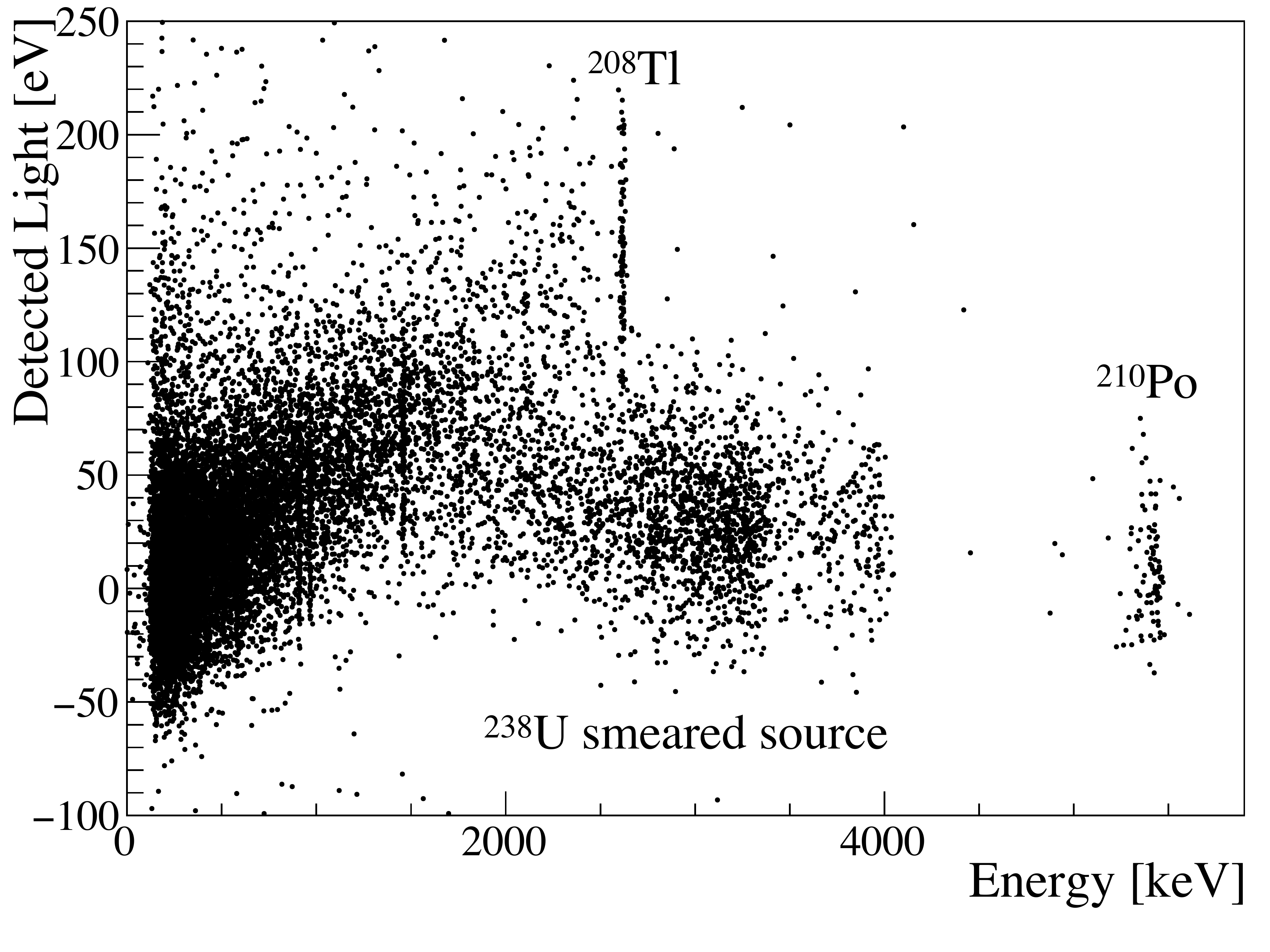}
\caption{Light vs heat scatter plot obtained in a 70 h measurement with the TeO$_2$ exposed 
to a $^{228}$Th source and a smeared $^{238}$U $\alpha$ source. Unfortunately $\alpha$ 
energy loss in the Mylar -constituting the smearing  medium- results in a tiny, but 
measurable, light emission that increases towards lower energies, i.e. at larger energy 
loss in the Mylar. The events above 4~MeV, on the contrary, are due to internal and/or 
surface contaminations and their light emission is compatible with zero 
(see text).}
\label{fig_4-scatter-plot}
\end{figure}

The light vs heat scatter plot is presented in Fig.~\ref{fig_4-scatter-plot} and shows an 
unexpected feature.
The $^{238}$U $\alpha$-events arising from the smeared source clearly show a tiny light 
emission that increases towards lower energies. This feature can only be ascribed to an 
energy loss in the  Mylar which emits few scintillation photons. To avoid this effect 
we usually face the aluminized surface of the Mylar towards the crystal so as to reflect the 
(very few) photons that could be produced in this plastic. This time however, 
we mistakenly mounted the Mylar with the uncoated side towards the detector.  This was confirmed after
subsequently opening the cryostat and checking.

The  result is shown in Fig.~\ref{fig_4-scatter-plot}: the amount of Cherenkov 
light, produced by a 2615~keV $\gamma$, that is collected with this new set-up is 
(151~$\pm$~4)~eV, 50 \% larger with respect to all our previous measurements with 
massive crystals~\cite{Casali-2017}, as well as roughly 50 \% larger  with respect to a 
measurement recently performed with a NTD-based light detector~\cite{Lumineu-2017} of the 
same type (considering the 40 \% reduced transmission area between BLD and crystal, as 
declared in the article).
The light distribution of  the 74 events belonging to the internal
$^{210}$Po $\alpha$ at 5407~keV (5304~keV  $\alpha$ + 103~keV nucleus recoil) shows a 
mean value of (5.8~$\pm$~3.3)~eV, still compatible with zero (see Sec.~\ref{sec:thermal_interference}) 
as it should be if the light only arises from the Cherenkov effect. More importantly, the width of 
the light distribution of $\alpha$'s is $\sigma_{\alpha}$=(22.7~$\pm$~ 2.7)~eV, fully compatible 
with the RMS noise of the BLD of Tab.~\ref{tab-new-BLD}.
The light signal induced by the 2615~keV $\gamma$ -on the contrary-  shows a width of 
$\sigma_{\gamma/\beta}$=(31.5~$\pm$~4.3)~eV   which is 
a result of the photostatistics and the light collection. 

In order to evaluate the Discrimination Power (DP) that can be obtained between the 
$\alpha$ and $\beta/\gamma$ distributions at 2528~keV (the Q$_{\beta\beta}$-value of 
the DBD  of $^{130}$Te) we use the same formula and arguments used 
in~\cite{Enriched-TeO2-Cherenkov-2017,Lumineu-2017}: the  DP can be quantified as the 
difference between the average values of the two distributions normalized to the square 
root of the quadratic sum of their  widths: 
\begin{equation} 
DP = \frac{|\mu_{\gamma/\beta}-\mu_{\alpha}|}{\sqrt{\sigma^{2}_{\gamma/\beta}+
\sigma^{2}_{\alpha}}}.
\label{eq:DP}
\end{equation} 
Re-scaling the light signal from 2615 to 2528~keV, we obtain DP=3.6, using one highly likely 
assumption that an $\alpha$ particle at 2528~keV will show a light signal equal than the same 
particle at~5304 keV ($^{210}$Po).
This DP is the best ever achieved with large mass TeO$_2$ crystals (M $>$ 7 g) and without 
the need for additional Neganov-Luke 
amplification~\cite{Lumineu-2017,Casali:2015gya,Gironi:2016nae}, or
more sophisticated TES sensors~\cite{Karo-2014} or both~\cite{Willers-2014}.

\section{Thermal conductance}\label{sec:thermal_interference}
As stated in Sec.~\ref{sec:Introduction}, the actual goal of this work was to experimentally 
demonstrate that the BLD can rest on the scintillating or luminescent crystal without heat sinking to it.  
Using the results in the previous section we can now calculate a limit on the heat flow through 
the Ge wafer and the TeO$_2$. If one assumes that a 5407 keV energy release in the TeO$_2$ produces
a mean value BLD signal that only depends on the heat flow (assuming no light emission), then we have 
an upper limit for the ratio of the heat flow through TeO$_2$ and Ge: 5.8~eV/5407~keV$\sim$10$^{-6}$.

In our case, an extremely low heat conductance was determined experimentally using static conditions. We measured 
the base resistance of the BLD  as 223.5 M$\Omega$ (corresponding to 11.8~mK), keeping the 
TeO$_2$ thermistor unbiased (i.e.  no power dissipation in it). We then gave the maximum  
(allowed by our biasing set-up) bias to the TeO$_2$ thermistor, corresponding to 
4.8 nA, and the TeO$_2$ thermistor changed its resistance from 626 M$\Omega$ 
(bias~$\rightarrow$~0) to 1.71 M$\Omega$.  The power dissipated on the TeO$_2$ was therefore 40 pW. 
The base resistance of the BLD decreased to 222.8 M$\Omega$, which corresponds to a temperature increase of 
only $\approx$~4.3 $\pm$ 0.2 ~$\mu$K. 
The same operation was performed with the BLD in working condition, i.e. bias current 
of 3.7 nA and a resistance of 1.47~M$\Omega$ (corresponding to $\sim$23~mK), and no variation
of the baseline of the BLD was registered.
A further investigation of the thermal conductance between a Ge-BLD and a TeO$_2$ crystal was performed
by exploiting a small TeO$_2$ crystal ($20~\times~20~\times~14$~mm$^{3}$, 34~g mass).
We used a standard BLD, i.e., the same thickness and height as in the previous discussion, but with the Ge wafer
held with PTFE clamps in a  stand-alone Cu mounting~\cite{NTD_LD_Lucifer-2013}.
For this experiment we rested the $20\times20$~mm$^{2}$ surface of the 34~g crystal 
 on the Ge wafer. The NTD thermistor-equipped TeO$_2$ crystal was surrounded  with the same reflecting 
foil and we performed the same measurement described in Sec.~\ref{scatter-section} with the same 
overall setup.
This time a 5304~keV $^{210}$Po decay occurring in the TeO$_2$ created a mean signal in the BLD 
of (317~$\pm$~29)~eV, definitively not compatible with the result of Sec.~\ref{scatter-section}.
The mean (light) signal registered in coincidence with the 2615~keV $\gamma$-line of $^{208}$Tl 
was (336~$\pm$~5)~eV. 
The $\alpha$-induced signal in the BLD, therefore, has to be ascribed to an effective thermal 
transfer from the TeO$_2$ to the BLD.
We can make a very rough estimation of the size of this transfer using the 
results of  the measurement of Sec.~\ref{scatter-section}.  If we assume the heat conductance 
to be linearly proportional to the pressure force  between the two mediums, then we may
simply compare the weight differences: 1.1 g in the case of the wafer resting onto the TeO$_2$ 
crystal versus 34~g in this last configuration.
Their ratio, i.e. 31, should be, in first approximation, the ratio between the thermal 
conductance in the two setups.  Ascribing the $\alpha$ signal of Sec.~\ref{scatter-section} 
exclusively to thermal transfer we would expect a thermal transfer signal of 
(180~$\pm$~90)~eV, which is compatible with the 317~eV observed during this measurement. On the other
hand, under the same assumption, we can evaluate the 2615-keV induced Cherenkov light signal
of this crystal as the difference between the observed signal and the re-scaled thermal transfer 
evaluated from  the $\alpha$. In this way we observe that the energy of the Cherenkov light 
emission in this 34~g  crystal is (185~$\pm$~15)~eV.

\section{Conclusions}
We have demonstrated the possibility of mounting BLDs by simply resting them on the surface of the 
corresponding scintillating crystal. With this new mounting method the light collection can increase up
to 50\% with respect to standard setups. We do not observe appreciable heat flow between the
scintillating crystal and BLD. 
We also improved the time response of our thermistor-based light detectors, reaching a rise
time of 0.8 ms and demonstrating that 0.5 ms is  achievable. This time response is necessary 
to remove the background induced by the pile-up of the 2$\nu$-DBD mode in the case 
of $^{100}$Mo-based crystals.  We reached a baseline resolution
of 20~eV RMS, more than 2 times better than the average value  our previous CUPID-0-like detectors. 
Thanks to these developments, we definitively demonstrated that standard thermistor-based
BLDs can be used for CUPID, both to read out the tiny Cherenkov light of TeO$_2$ as well as to
read out the Mo-based scintillating crystals.  

We do believe that this simplified technique could be applied to any kind of BLD, irrespective
of the sensor type. The first approximation thermal conductance between crystal and BLD 
does not depend upon the energy of the phonons, so we would expect that thermal transfer 
would be as negligible in TES or MMC devices as it is in our NTDs.  
More generally this new technique could be also  applied  in  the case of stacked, standard small
bolometers, provided that the weight does not exceed a
few grams. However, since the measured thermal transfer is rather small, the weight of the
bolometer will not be a significant limiting factor in low energy threshold applications.  

\section{Acknowledgments}
This work was performed within the CUPID experiment founded by INFN and supported by
the National Science Foundation under Grant NSF-PHY-1614611.

We thank the CUPID-0 and the CUORE collaborations for the overall support and for sharing their 
DAQ and software.
We  express our gratitude to LNGS for the generous hospitality  and, in particular, to
the mechanical workshop personnel including E. Tatananni, A. Rotilio, A. Corsi, and B.
Romualdi for their continuous and  constructive help. We are  also grateful to  M. Guetti 
for his invaluable support and expertise  in the cryostat facility maintenance. 
We acknowledge Dr. C. Arnaboldi for his precious 
support, even though he has left this field of research many years ago.  We are especially 
grateful to E. Ferri for her kind support in the thermistor wire-bonding.


\end{document}